\documentclass[twocolumn,showpacs,preprintnumbers,aps]{revtex4-1}

\usepackage{graphicx,color}

\usepackage{dcolumn}

\usepackage{bm}

\begin{document}

\preprint{}

\title{Pressure dependence of the magnetic ground state in CePtSi$_2$}

\author{S. E. Dissanayake}
\altaffiliation[Present address: ]{Department of Mechanical Engineering, University of Rochester, Rochester, New York 14627, USA}
\affiliation{Neutron Scattering Division, Oak Ridge National Laboratory, Oak Ridge, Tennessee 37831, USA}

\author{H. Muto}
\affiliation{Graduate School of Science and Technology, Niigata University, Ikarashi, Niigata 950-2181, Japan}

\author{S. Suzuki}
\affiliation{Graduate School of Science and Technology, Niigata University, Ikarashi, Niigata 950-2181, Japan}

\author{T. Nakano}
\affiliation{Graduate School of Science and Technology, Niigata University, Ikarashi, Niigata 950-2181, Japan}

\author{S. Watanabe}
\affiliation{Department of Basic Sciences, Kyushu Institute of Technology, Kitakyushu, Fukuoka 804-8550, Japan}

\author{F. Ye}
\affiliation{Neutron Scattering Division, Oak Ridge National Laboratory, Oak Ridge, Tennessee 37831, USA}

\author{W. Tian}
\affiliation{Neutron Scattering Division, Oak Ridge National Laboratory, Oak Ridge, Tennessee 37831, USA}

\author{J. Gouchi}
\affiliation{Institute for Solid State Physics, University of Tokyo, Kashiwa, Chiba 277-8581, Japan}

\author{M. Matsuda}
\email[Corresponding author: ]{matsudam@ornl.gov}
\affiliation{Neutron Scattering Division, Oak Ridge National Laboratory, Oak Ridge, Tennessee 37831, USA}

\author{Y. Uwatoko}
\affiliation{Institute for Solid State Physics, University of Tokyo, Kashiwa, Chiba 277-8581, Japan}

\date{\today}

\begin{abstract}
CePtSi$_2$ was reported to exhibit an antiferromagnetic order below $T\rm^*$=1.8 K at ambient pressure, a valence state change at $\sim$1.2 GPa, and superconductivity in the range between 1.4 and 2.1 GPa with the maximum transition temperature of 0.14 K [T. Nakano $et$ $al$., Phys. Rev. B 79, 172507 (2009)]. We have performed polycrystalline and single crystal neutron diffraction experiments to determine the magnetic structure under ambient and high pressures. We found that incommensurate magnetic peaks with the magnetic propagation vector of (0.32, 0, 0.11) at ambient pressure below $T\rm_{SDW}\sim$1.25 K. Those magnetic peaks which originate from a spin-density-wave order with the easy axis along the $c$ axis and an averaged ordered moment of 0.45(5)$\mu\rm_B$, suggesting that there may be an intermediate phase between $T\rm^*$ and $T\rm_{SDW}$. Applying pressures, the magnetic propagation vector shows no change and the magnetic order disappears around 1.0 GPa, which is much lower than the critical pressure for the superconducting phase.
The results suggest that other than magnetic fluctuations may play a primary role in the superconducting pairing mechanism.
\end{abstract}


\maketitle

\section{Introduction}
Unconventional spin-mediated superconductor, which is discussed widely in cuprates, iron based systems, and heavy fermionic systems, has attracted considerable attention. Recently, there were reports of a new mechanism of superconductivity: ``valence-mediated superconductivity", a novel mechanism \cite{Miyake1999,Onishi2000,Watanabe2007,Holmes2007} that
would be the third one after the phonon- and spin-mediated superconductivity. It was proposed that the higher pressure superconducting phase in CeCu$_2$(Si,Ge)$_2$ originates from the valence fluctuations \cite{Watanabe2006,Watanabe2011}.
A sharp change in the nuclear-quadrupole-resonance (NQR) frequency observed under pressure $P\rm_v=4.5\pm 0.2$~GPa in CeCu$_2$Si$_2$ evidenced the valence crossover, which supports the valence-fluctuation mediated superconductivity \cite{Kobayashi2013}. Around $P=P\rm_v$, the residual resistivity has a peak under pressure, where the resistivity shows the remarkable non-Fermi liquid behavior $\rho(T)\sim T$ \cite{Holmes2004}. These behaviors are also explained by the enhanced valence fluctuations of the Ce ions \cite{Miyake1999,Holmes2004}.

\begin{figure}
\includegraphics[width=8.5cm]{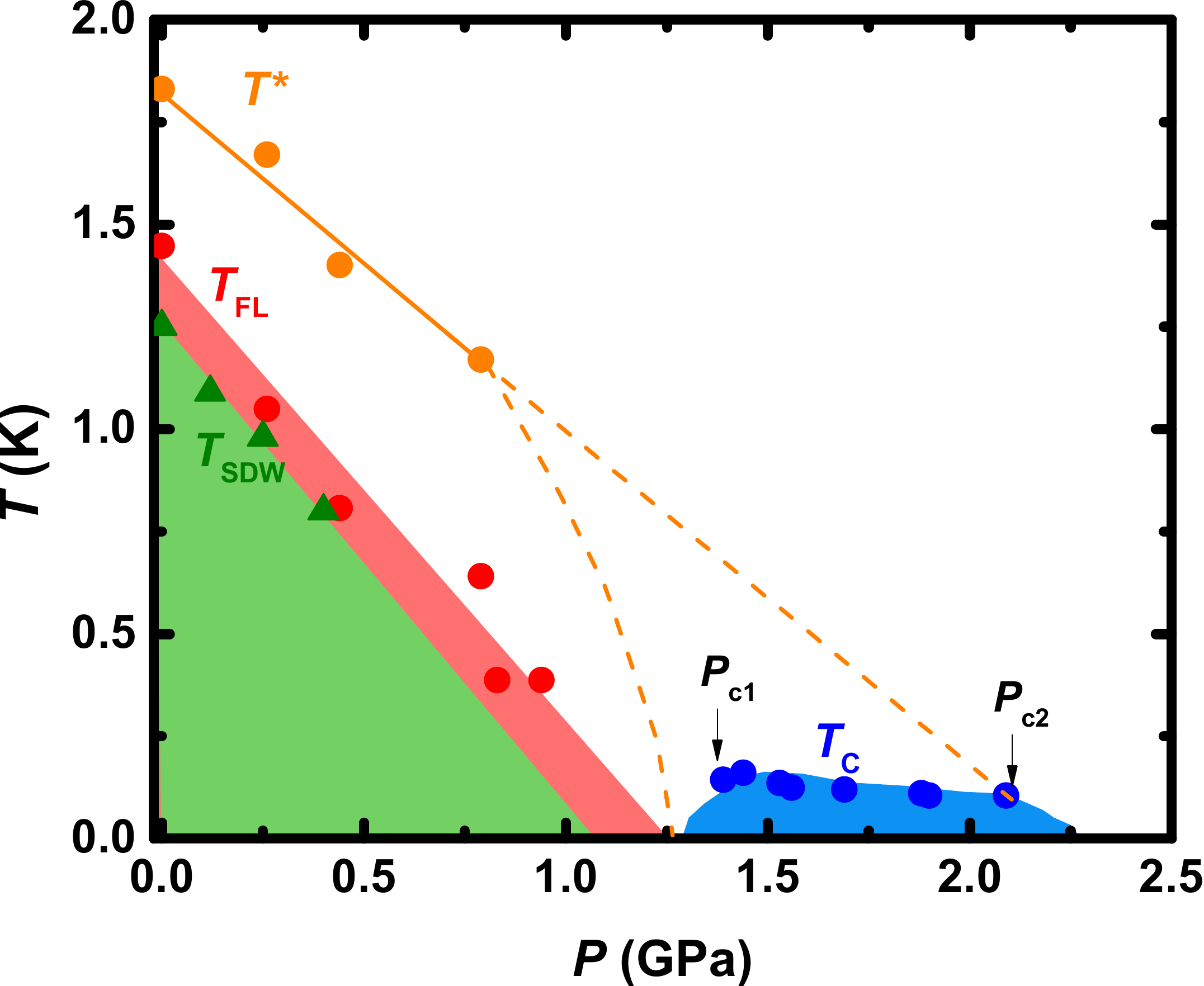}
\caption{Temperature-pressure phase diagram of CePtSi$_2$.
The filled circles and triangles represent the data in Refs. \onlinecite{Nakano2009}, \onlinecite{Nakano2010}, and \onlinecite{Nakano2013} and the present results, respectively. The lines are guides to the eye.}
\label{PD}
\end{figure}

CePtSi$_2$ has the CeNiGe$_2$-type orthorhombic layered structure. It shows a large decrease of resistivity below $\sim$1.8 K \cite{Nakano2009} and a peak of heat capacity at the same temperature \cite{Hirose2011} under ambient pressure, which was ascribed to an antiferromagnetic ordering temperature ($T^{*}$) \cite{TN}. There is another anomaly of resistivity at a lower temperature of $T\rm_{FL}$ ($\sim$1.4 K), below which the resistivity shows the Fermi liquid behavior with $T^2$-dependence \cite{Nakano2010,Nakano2013}. Applying pressure, both $T^{*}$ and $T\rm_{FL}$ decrease and superconductivity appears at $P\rm_{c1}$= 1.4 GPa \cite{Nakano2009,Hirose2011}, as shown in Fig. \ref{PD}. The maximum superconducting transition temperature is 0.14 K around 1.4 GPa.
The resistivity of CePtSi$_2$ shows two local maxima $T_1$ (4 K) and $T_2$ (33 K) at ambient pressure \cite{Nakano2009,Nakano2010,Nakano2013}. These two maxima are well-known characteristics of Ce based Kondo compounds, showing an interplay between the Kondo effect and crystalline electric field (CEF) splitting, as reported in CeCu$_2$Ge$_2$ \cite{Jaccard1999}.
Applying pressure, $T_2$ is almost unchanged, whereas $T_1$ starts to increase above $P\rm_{v}$ ($\sim$1.2 GPa) and gradually approach $T_2$, suggesting that a valence change occurs around $P\rm_v$ as a crossover in CePtSi$_2$ \cite{Nakano2013}.
The resistivity data also show the quantum critical behavior around $\sim$1.2 GPa, where non-Fermi liquid behavior was observed \cite{Nakano2009,Nakano2013}.
Around $P\rm_v(\sim1.2$~GPa), the residual resistivity has a peak and the $T$-linear-like resistivity was observed in CePtSi$_2$ \cite{Nakano2009,Nakano2013}.
Since $P\rm_{v}$ is close to $P\rm_{c1}$, it is expected that the valence fluctuation mediates the superconductivity in this material \cite{Nakano2009,Nakano2010,Nakano2013}.
Up to date, the detailed magnetic structure and its pressure dependence in CePtSi$_2$ have not been clarified, which is critical to understand the nature for the quantum critical behavior and the superconductivity.

We performed a neutron diffraction study on the magnetic structure of CePtSi$_2$ as a function of pressure. We found that the magnetic structure at ambient pressure is a spin-density-wave (SDW) structure with the magnetic propagation vector of (0.32, 0, 0.11). The magnetic transition temperature ($T\rm_{SDW}$) was clearly determined to be 1.25(3) K, which is much lower than $T\rm^*$ previously determined from the resistivity and heat capacity measurements and is rather close to $T\rm_{FL}$. Applying pressure, the magnetic propagation vector is pressure independent and the magnetic order almost disappears around 1.0 GPa, which is much lower than $P\rm_{c1}$.
These results suggest that magnetic fluctuations may not play a primary role in the superconducting pairing mechanism in CePtSi$_2$, which should be elucidated with further studies.

\section{Experimental details}
A polycrystalline sample of CePtSi$_2$, used for measurements at ambient pressure, was synthesized by arc-melting stoichiometric amounts of Ce (3N), Pt (4N), and Si (5N) in an Ar atmosphere with additional remelting and post-annealing to ensure homogeneity \cite{Nakano2009}. Single crystal samples, used for measurements at ambient and high pressures, were grown by the Czochralski method in a tetra-arc furnace, using the CePt$_{1.1}$Si$_{2.2}$ ingot, after the previously reported method \cite{Kasaya1996}. The bulk properties of the single crystal samples are consistent with those reported previously \cite{Nakano2009,Hirose2011,Nakano2010,Nakano2013}, as shown in the Supplemental Material \cite{SupplMater}.
Note that the crystals grown using different methods show very similar bulk properties, including the transition temperatures. This suggests that the sample dependence is negligibly small.

High pressure single crystal neutron diffraction measurements were performed using the time-of-flight diffractometer CORELLI \cite{corelli} at the Spallation Neutron Source (SNS) and the triple-axis spectrometers HB-1 and HB-1A at the High Flux Isotope Reactor (HFIR) at Oak Ridge National Laboratory (ORNL). The hydrostatic pressures were generated with a self-clamped piston-cylinder cell (SCPCC) made of a Zr-based amorphous alloy \cite{cell}. The crystal dimensions were 1.2$\times$1.2$\times$3 mm$^{3}$. Fluorinert was chosen as the pressure transmitting medium. The pressure inside SCPCC was monitored by measuring the lattice constant of a comounted NaCl crystal. We found that the pressure is reduced by a few percent on cooling from room temperature to 0.3 K.
A $^3$He refriegerator was used to cool down the polycrystalline and single crystal samples down to 0.3 K. The single crystal was mounted with $(H0L)$ in the horizontal scattering plane.

For magnetic structure analysis, the representation analysis was performed using the SARAh package~\cite{sarah}. Rietveld refinements were performed for polycrystalline and single crystal diffraction data using the FullProf package~\cite{fullprof}. The magnetic form factor for Ce$^{3+}$ was used for the magnetic structure refinement. This is considered to be reasonable since the primary purpose for the refinement is to distinguish between two possible magnetic structure models, as described in Sec. IIIA.

\section{Experimental Results}
\subsection{Magnetic structure at ambient pressure}
\begin{figure*}
\includegraphics[width=15cm]{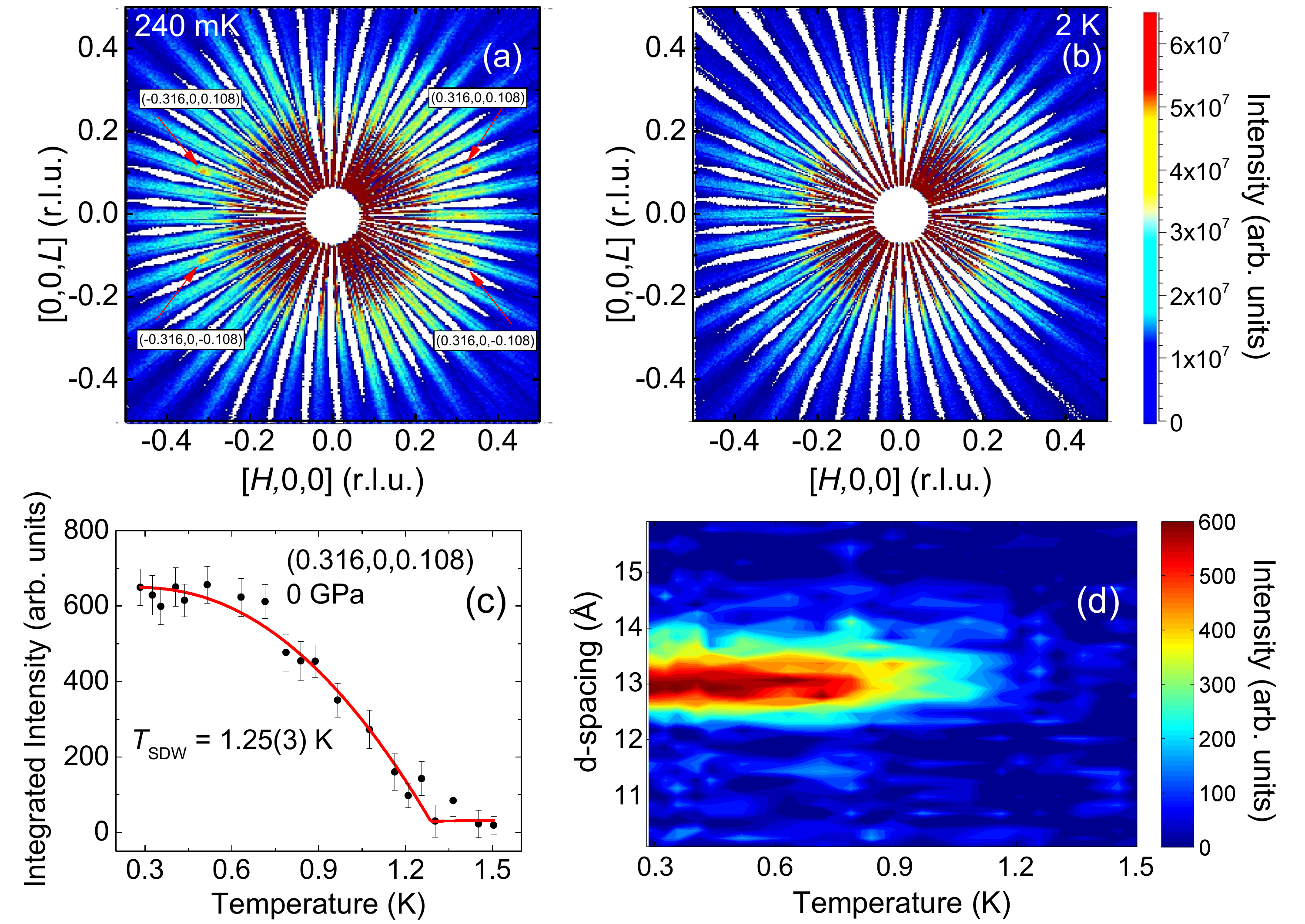}
\caption{Contour maps of neutron diffraction intensity at a low-$Q$ region in the $(H0L)$ plane in CePtSi$_2$ measured at 0.24 K (a) and 2.0 K (b) at ambient pressure. Magnetic peaks are observed at (0.32, 0, 0.11), ($-$0.32, 0, 0.11), (0.32, 0, $-$0.11), and ($-$0.32, 0, $-$0.11) at 0.24 K. (c) and (d) Temperature dependence of the (0.32, 0, 0.11) magnetic Bragg peak intensity. The solid line is the result of a fit to a power-law function.}
\label{CORELLI}
\end{figure*}
In order to search for magnetic signal in CePtSi$_2$, we first preformed a neutron diffraction measurement using a polycrystalline sample. However, no magnetic signal was observed down to 0.3 K. Then, a single crystal was measured on a time-of-flight diffractometer CORELLI, which has wide-range two-dimensional detectors suitable for observing incommensurate Bragg peaks. As shown in Figs. \ref{CORELLI}(a) and \ref{CORELLI}(b), a magnetic Bragg peak was observed at (0.32, 0, 0.11) and its three equivalent positions at 0.24 K, whereas those peaks disappear at 2 K. Figure \ref{CORELLI}(c) shows the magnetic intensity at (0.32, 0, 0.11) as a function of temperature. The magnetic intensity develops below 1.25(3) K, where the low temperature magnetic phase was expected. No other magnetic signal was observed between 1.25 and 1.8 K, as shown in Fig. \ref{CORELLI}(d), where an intermediate magnetic phase was predicted. We tried to find other magnetic Bragg peaks below 1.25 K other than the four peaks in order to perform a magnetic structure analysis. However, no additional peaks were found. As described below, this is because the magnetic peaks at higher-$Q$'s were not covered by the vertical detector range.

\begin{figure}
\includegraphics[width=9.0cm]{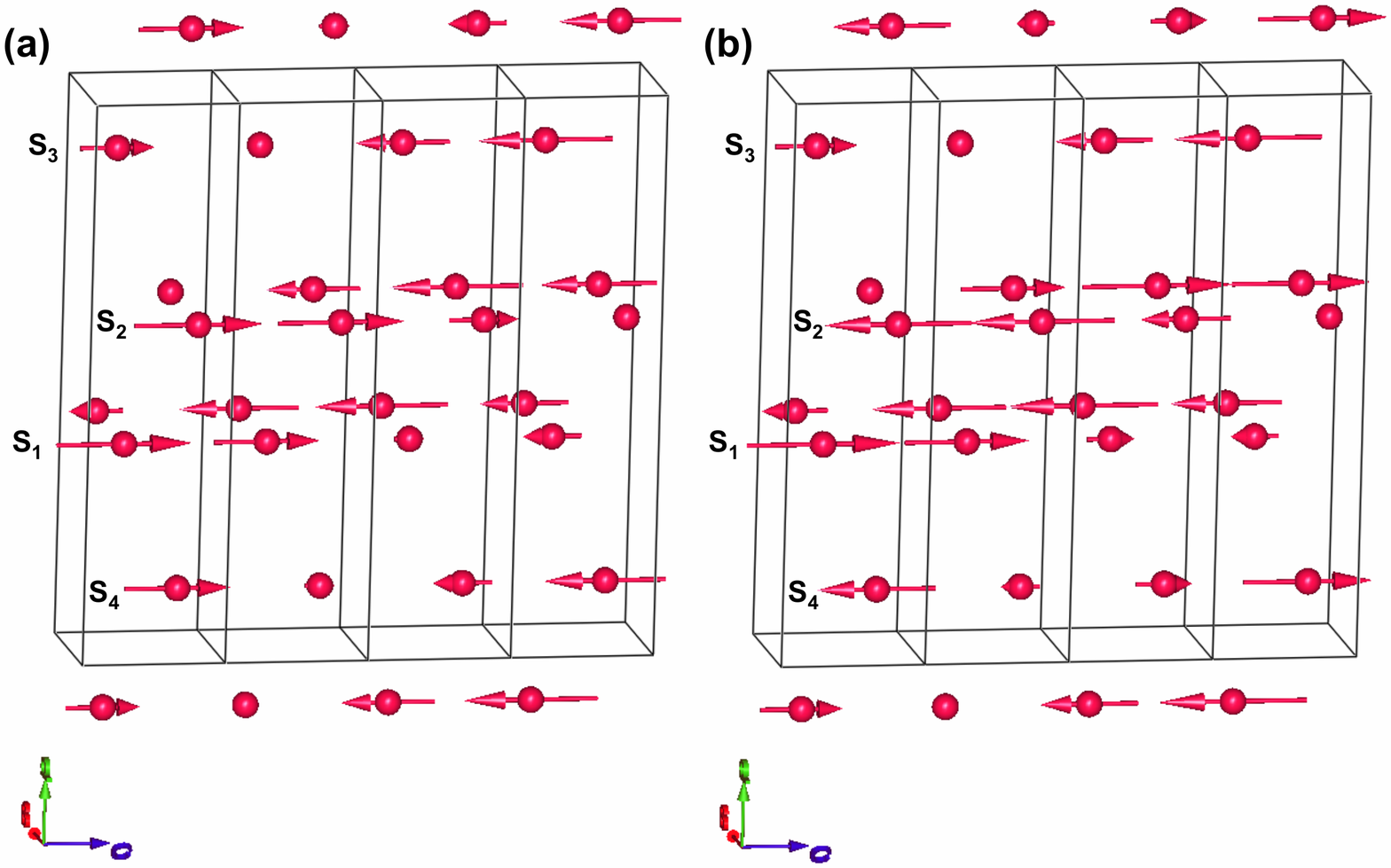}
\caption{Two possible spin density wave structures for CePtSi$_2$. The spin directions of $S_1$ and $S_2$ as well as those of $S_3$ and $S_4$ are parallel in Model 1 (a) and antiparallel in Model 2 (b), respectively. Model 1 was found to be the appropriate magnetic structure.
Atomic coordinates of the atoms corresponding to $S_1$, $S_2$, $S_3$ and $S_4$ are, $S_1$: $( 0,~ 0.39465,~ 0.25)$, $S_2$: $( 0,~ 0.60535,~ 0.75)$, $S_3$: $( 0.5,~ 0.89465,~ 0.25)$ and $S_4$: $( 0.5,~ 0.10535,~ 0.75)$. (See the Supplementary Material \cite{SupplMater}.)}
\label{Structures}
\end{figure}
\begin{figure}
\includegraphics[width=8.8cm]{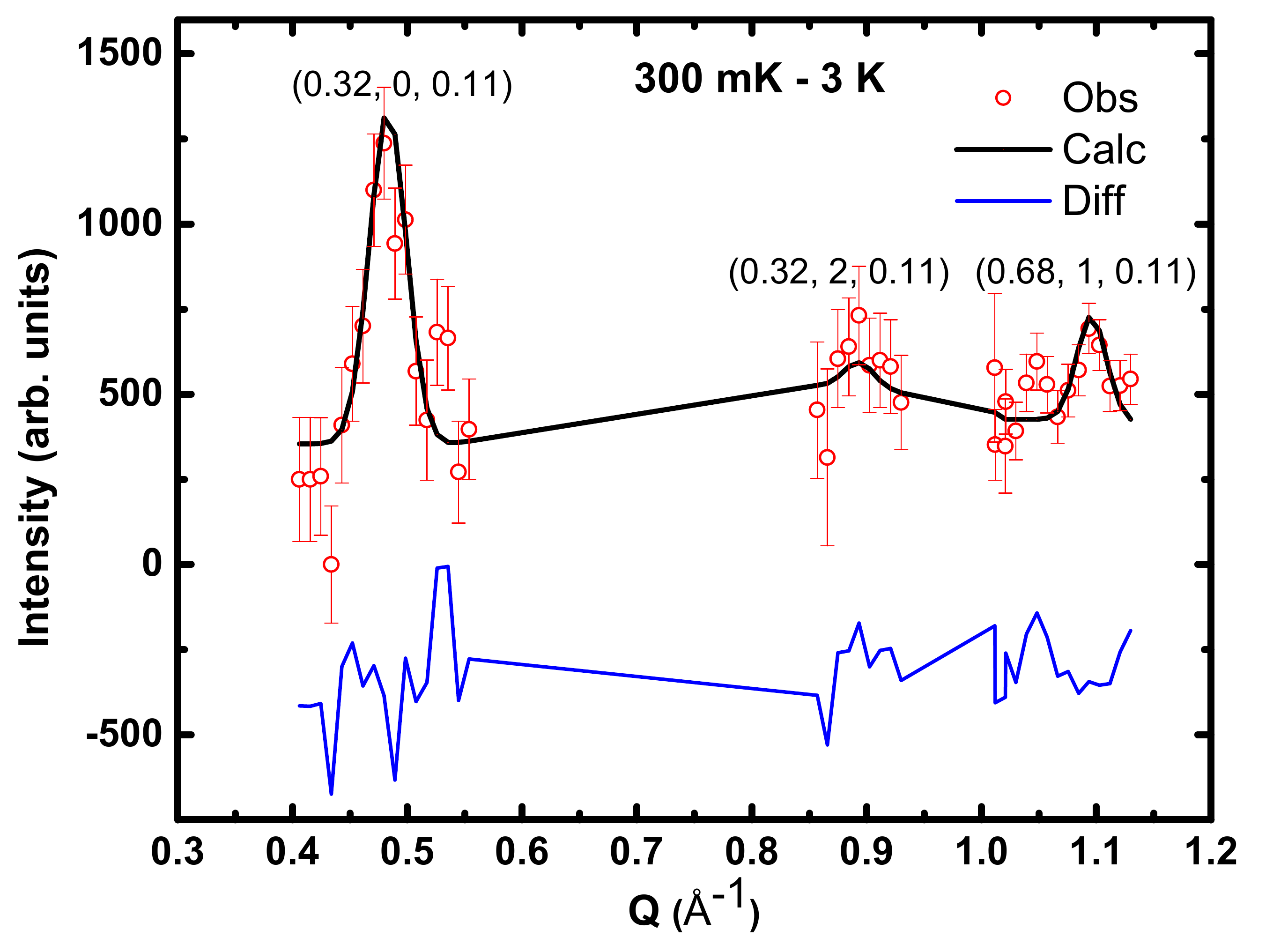}
\caption{Neutron diffraction intensities around (0.32, 0, 0.11), (0.32, 2, 0.11), and (0.68, 1, 0.11) measured on HB-1A using polycrystalline sample of CePtSi$_2$. The background signal measured at 3 K was subtracted. Constant value was added to help the refinement using Fullprof. The open circles are observed points. The bold solid line represents the result of the Rietveld refinement. The thin blue line is the difference between the observed and fitted intensities.
The magnetic form factor for Ce$^{3+}$ was used for the magnetic structure refinement. The residual intensity at $Q\sim$1.05 \AA$^{-1}$ is too sharp for actual magnetic peak and is most probably due to imperfect background subtraction.}
\label{HB1A}
\end{figure}
We performed a representation analysis \cite{Bertaut62,Bertaut68,Bertaut71,Bertaut81,Izyumov-book,Bradley_Cracknell,Cracknell,Wills-J-de-Physique,Wills-PRB-Jarosites,Kovalev} to narrow down the possible magnetic structures below 1.25 K, given that a magnetic Bragg peak was observed at (0.32, 0, 0.11). There are two candidates for the magnetic structure, as shown in Fig. \ref{Structures}. Model 1 and Model 2 are obtained using  basis vectors $\bf {\psi_{6}}$ from $\Gamma_{2}$ and $\bf {\psi_{3}}$ from $\Gamma_{1}$, respectively. (See the Supplementary Material \cite{SupplMater}.) Both Models 1 and 2 are spin density wave type structures with an easy axis along the $c$ axis. The spin directions of $S_1$ and $S_2$ as well as those of $S_3$ and $S_4$, where $S_i$ ($i$=1, 2, 3, and 4) are four Ce moments in a unit cell \cite{SupplMater}, are parallel in Model 1 and antiparallel in Model 2, respectively.
In order to distinguish the two magnetic structure models, the magnetic intensities at (0.32, 2, 0.11) and (0.68, 1, 0.11) should be evaluated, which were not measurable on CORELLI using the single crystal because those peaks are out of the vertical detector coverage.
In the Model 1, it is expected that the (0.32, 2, 0.11) intensity is negligibly weak and the (0.68, 1, 0.11) intensity is a factor of $\sim$4 weaker than the (0.32, 0, 0.11) intensity. On the other hand, in the Model 2, both the (0.32, 2, 0.11) and (0.68, 1, 0.11) intensities are expected to be larger than the (0.32, 0, 0.11) intensity.
\begin{figure*}
\includegraphics[scale=0.565]{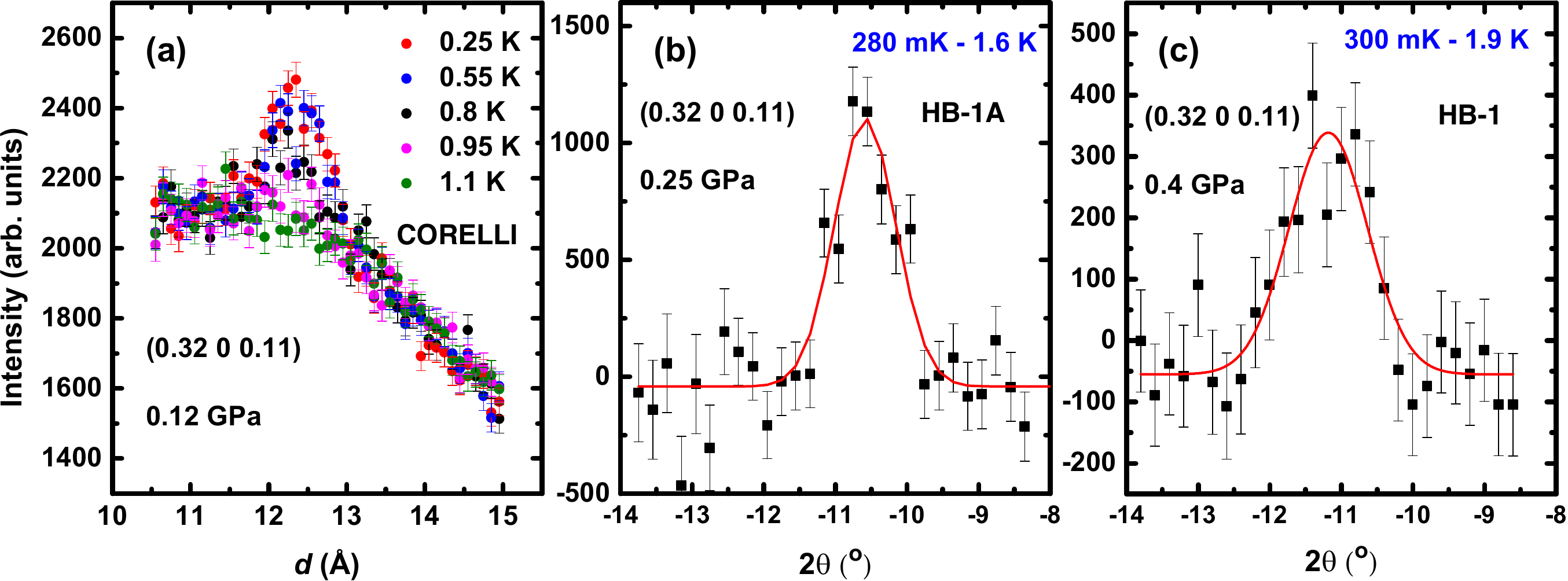}
\caption{(0.32, 0, 0.11) magnetic Bragg peak measured at 0.12 GPa (a), 0.25 GPa (b), and 0.4 GPa (c). The background intensities measured above $T\rm_{SDW}$ are subtracted in (b) and (c). The solid lines are the results of fits to a Gaussian function.
The data shown in (a),(b) and (c) were measured at CORELLI diffractometer, HB-1, and HB-1A triple axis spectrometers, respectively. (a) represents $d$-spacing dependence of integrated intensities obtained using the instrument view window in CORELLI, while (b) and (c) represent background subtracted $\theta$-2$\theta$ scans.}
\label{Pressure1}
\end{figure*}
\begin{figure*}
\includegraphics[scale=0.565]{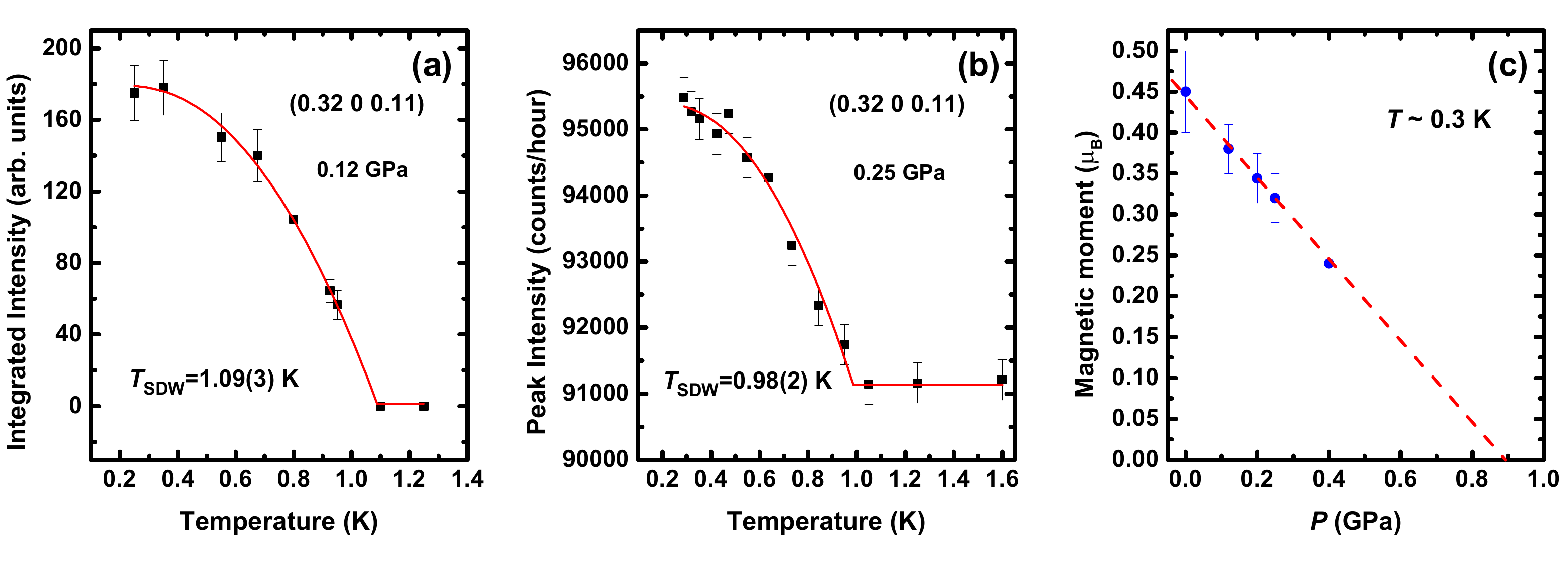}
\caption{Temperature dependence of the (0.32, 0, 0.11) magnetic Bragg peak intensity measured at 0.12 GPa (a) and 0.25 GPa (b). The solid lines are the fitting results to a power-law function. (c) Averaged magnetic moment around 0.3 K as a function of applied pressure. The broken line is a guide to the eye.}
\label{Pressure2}
\end{figure*}

In order to observe the magnetic Bragg intensities at the mentioned (0.32, 2, 0.11) and (0.68, 1, 0.11) reflections, we performed a neutron diffraction measurement using a polycrystalline sample on HB-1A. Figure \ref{HB1A} shows the neutron diffraction pattern around (0.32, 0, 0.11), (0.32, 2, 0.11), and (0.68, 1, 0.11) observed at 0.3 K \cite{note}.
The magnetic Bragg intensities at (0.32, 2, 0.11) and (0.68, 1, 0.11) are much weaker than that at (0.32, 0, 0.11), revealing that the Model 1 is the appropriate magnetic structure at ambient pressure.
The averaged magnetic moment was determined to be 0.45(5) $\mu_B$. It is worth noting that this magnetic structure is the simplest magnetic structure that can be derived from the available data. More complicated structure or small deviations from the above structure may be possible if further weak magnetic peaks could be detected.

\subsection{Pressure dependence of the magnetic ground state}
The high pressure neutron diffraction measurements using single crystal samples were performed on CORELLI, HB-1A, and HB-1.
Since the magnetic moment is small ($<$0.5$\mu\rm_B$), the measurements were challenging due to high background and low beam transmission originating from the pressure cell and pressure transmitting medium. The Bragg peak intensities around (0.32, 0, 0.11) at 0.12, 0.25, and 0.4 GPa are plotted in Fig. \ref{Pressure1}. We found that the magnetic propagation vector is almost independent of pressure. 

The order parameters of the integrated magnetic intensities at 0.12 and 0.25 GPa are plotted in Figs. \ref{Pressure2}(a) and \ref{Pressure2}(b), respectively. $T\rm_{SDW}$ gradually decreases with increasing pressure, as shown in Fig. \ref{PD}. The averaged magnetic moment as a function of pressure is shown in Fig. \ref{Pressure2}(c). The magnetic moment also decreases gradually with increasing pressure. The magnetic intensity at 0.4 GPa, where magnetic moment is 0.24(3)$\mu\rm_B$, is already very weak and reaches the limit for measuring magnetic signal with reasonably good signal-to-noise ratio. Therefore, we did not measure at higher pressures than 0.4 GPa in this study. As summarized in Fig. \ref{PD}, $T\rm_{SDW}$ is systematically lower than the $T^{*}$, obtained from the resistivity measurements. Extrapolating the $T\rm_{SDW}$-pressure (Fig. \ref{PD}) and magnetic moment-pressure [Fig. \ref{Pressure2}(c)] relations to higher pressures, the long-range SDW order is expected to disappear at $\sim$1.1$\pm0.15$ and $\sim$0.9$\pm0.1$ GPa, respectively.
\section{Discussion}
As described in Sec. I, the resistivity becomes $T$-linear-like and the residual resistivity shows a maximum at the quantum critical pressure $\sim$1.2 GPa \cite{Nakano2009}. This pressure corresponds to the pressure where a valence crossover or a valence transition was reported \cite{Nakano2013}, suggesting that the quantum critical behavior is driven by the valence fluctuations in CePtSi$_2$.
Our results suggest that the long-range magnetic order may disappear around $\sim$1.0 GPa, which is lower than the quantum critical pressure ($\sim$1.2 GPa) and $P\rm_{c1}$($\sim$1.4 GPa). This implies that the magnetic critical point may not be directly related with the quantum critical behavior and the appearance of superconductivity.

The temperature-pressure phase diagram in Ce- and Yb-based heavy fermion metals with valence fluctuations was discussed in Ref. \cite{Watanabe2011}. The interplay of the magnetic order and valence fluctuations is important to understand the physical properties in the Ce- and Yb-based systems. Applying pressure, valence fluctuations are gradually enhanced, which leads to the suppression of a magnetic order. $P\rm_m$ and $P\rm_v$ are defined as critical pressures, where the magnetic order is suppressed completely and the valence transition occurs, respectively.  There are three categories with (a) $P\rm_m$ $<$ $P\rm_v$, (b) $P\rm_m$ = $P\rm_v$, and (c) $P\rm_m$ $>$ $P\rm_v$. The phase diagram changes from (a) to (c) with reducing the $c$-$f$ mixing in the periodic Anderson model. The phase diagram in CePtSi$_2$  is considered to be located between (a) and (b) and close to (b), since $P\rm_v$ is reported to be $\sim$1.2 GPa \cite{Nakano2009} and the present results suggest that $P\rm_m\sim$1.0(1) GPa. The superconducting phase is located between 1.4 and 2.1 GPa, which is higher than $P\rm_m$ and $P\rm_v$, where the superconducting transition temperature is supposed to be enhanced.
Although the superconducting phase was shown to be stabilized in the extended region around $P=P\rm_v$ \cite{Onishi2000,Watanabe2006}, the quantitative explanation for the measured superconducting region should be addressed by future theoretical studies. 

One of the most remarkable findings in this study is that $T\rm_{SDW}$ obtained from neutron diffraction measurements is much lower than $T^{*}$ determined by resistivity and heat capacity measurements \cite{Nakano2009,Hirose2011}.
As described in Sec. I, an additional anomalous temperature ($T\rm_{FL}$) was reported below $T^{*}$. The transition temperature and its pressure dependence of $T\rm_{FL}$ are similar to those of $T\rm_{SDW}$ (Fig. \ref{PD}), suggesting that the $T\rm_{FL}$ corresponds to $T\rm_{SDW}$.
Furthermore, the heat capacity shows a broad shoulder around 1.25 K \cite{Hirose2011,SupplMater}. This may also correspond to the antiferromagnetic transition. This is in contrast to CeRhGe$_2$ which shows a sharp lambda transition \cite{Hirose2011}.
These results suggest that an additional phase exists between $T^{*}$ and $T\rm_{SDW}$/$T\rm_{FL}$ and the magnetic state may be disordered in the phase, since we did not observe magnetic Bragg peaks.
It is puzzling why no sharp anomaly is observed at $T\rm_{SDW}$ in the resistivity and heat capacity measurements.

Here, we discuss a possible state in the intermediate phase between $T^{*}$ and $T\rm_{FL}$/$T\rm_{SDW}$. 
Since no magnetic Bragg peaks were observed in this phase, a long-range magnetic order does not likely occur.
One possible state is a quadrupole order. The CEF ground state in CePtSi$_2$ is reported as $|\psi_{\pm}\rangle=0.656|\pm 3/2\rangle+0.288|\mp 1/2\rangle+0.698|\mp 5/2\rangle$ \cite{Hirose2011}. Then, it is possible that the electric quadrupole originates from the transition between the $J_z=\pm 3/2$ and $J_z=\mp 1/2$ states and/or the $J_z=\mp 1/2$ and $J_z=\mp 5/2$ states.
As discussed in Ref. \cite{Watanabe2011}, the magnetic order can be suppressed by valence fluctuations enhanced at the valence-crossover pressure or by valence-transition pressure $P\rm_v$. In the CEF of the present system, the electric quadrupole order can occur as mentioned above. In this case, the quadrupole order is also considered to be suppressed by $P\rm_v$.
The Ce site is not centro-symmetric locally, which gives rise to the odd-parity CEF. Therefore, in this case, an odd-parity multipolar order can be induced. In $\beta$-YbAlB$_4$, magnetic toroidal degree of freedom is induced under the odd-parity CEF \cite{Watanabe2019}.
If the odd-parity multipole is composed of 4$f$ and 5$d$ orbitals at Ce, the origin of the ordering of the multipole and the emergence of the valence-crossover pressure or valence-transition pressure $P\rm_v$ is common, which is the Coulomb repulsion between the 4$f$ and 5$d$ orbitals at Ce \cite{Watanabe2019}.

\section{Summary}
Our neutron diffraction study in CePtSi$_2$ has revealed that the magnetic structure is SDW with the magnetic propagation vector of (0.32, 0, 0.11) and the easy axis along the $c$ axis at ambient pressure. $T\rm_{SDW}$ ($\sim$1.25 K) is much lower than $T\rm^*$ ($\sim$1.8 K) but close to $T\rm_{FL}$ ($\sim$1.4 K), suggesting that there may be an intermediate phase between $T\rm^*$ and $T\rm_{SDW}$, which might be a quadruple or odd-parity multipoler ordered state.
Applying pressure,
the magnetic order may disappear around 1.0 GPa, which is lower than $P\rm_{c1}$($\sim$1.4 GPa), suggesting that magnetic fluctuations are not directly coupled to the superconducting pairing mechanism.
Further experimental and theoretical studies are highly desirable to clarify the superconducting pairing mechanism in CePtSi$_2$.

\section*{Acknowledgments}
We thank R. Kobayashi for his preliminary measurement at the early stage of this research. This research was supported in part by the U.S.-Japan Cooperative Program on Neutron Scattering. This research used resource at the Spallation Neutron Source and High Flux Isotope Reactor, DOE Office of Science User Facilities operated by the Oak Ridge National Laboratory. This work was partially supported by MEXT, the Grant-in-Aid for Scientific Research ``Grant No. 19H00648''.


\begin{thebibliography}{}
\bibitem{TN}$T\rm_N$ was used to represent this characteristic temperature in the literature. $T^*$ is used in this paper since no magnetic order was observed below the temperature.
\bibitem{note}Note that the magnetic Bragg peaks were not observed in the preliminary powder diffraction measurement on a neutron powder diffractometer, because the intrinsic high instrumental $Q$ resolution is compensated by the low beam flux, that is not strong enough to observe the weak magnetic signals.
\end{thebibliography}
\end{document}